\documentstyle[12pt]{article}

\author{S. S. Safonov}
\title{Tomography of Quantum Spinor States}
\date{}
\begin{document}

\begin{center}
{\bf \Huge Tomography of Quantum Spinor States\\}
\end{center}

\vspace*{1cm}

\begin{center}
{\bf \Large S. S. Safonov\\}
\end{center}

\begin{center}
{\it Moscow Institute of Physics and Technology,}\\
{\it Institutskii pr. 9, Dolgoprudny, Moscow reg., 141700, Russia\\
and\\
P. N. Lebedev Physical Institute, Russian Academy of Science,\\
Leninskii pr. 53, Moscow 117924, Russia} 
\end{center}

\begin{abstract}
A possibility of describing two-level atom states in terms of positive 
probability distributions (analog to the symplectic tomography scheme) 
is considered. As a result the basis of the irreducible representation 
of a rotation group can be realized by a family of the probability 
distributions of the spin projection parametrized by points on the
sphere. Furthermore the tomography of rotational states of molecules and 
nuclei which can be described by the model of a symmetric top is discussed. 
\end{abstract}

\vspace*{1cm}

Recently \cite{Man'ko1,Man'ko2} the symplectic tomography scheme was suggested 
to obtain the Wigner function by measuring the probability distribution function 
(the marginal distribution) for a shifted, rotated and squeezed quadrature, 
which depends on extra parameters. The invertable map of the Wigner function 
of a quantum state onto the positive marginal distribution for the continous 
observable (coordinate) was used to give the formulation of quantum dynamics
as the classical statistical process \cite{Man'ko3,Man'ko4,Man'ko5}. From this 
point of view, the Moyal approach \cite{Moyal} to the quantum evolution as to 
a statistical process was improved in the sense that instead of the Moyal 
quasidistribution function (the Wigner function) was introduced the positive 
probability distribution of the measurable variables describing an arbitrary 
quantum state and its evolution. A spin states reconstruction procedure similar 
to the symplectic tomography scheme is investigated in \cite{Dodonov,Olga1}. The 
authors of these papers showed that in the framework of the symplectic
tomography approach it is possible to describe equivalently the spin state in 
terms of the classical distribution function of a discrete variable instead of
the wave function or the density matrix. Taking into account the results of 
these papers in this article it will be investigated the tomography scheme for 
two-level atom states and rotational states of molecules and nuclei.

The goal of this work is to make a review of construction of the explicit 
formula for the invertable map connecting the quantum state of a two-level 
atom described by the $1/2$-spin density matrix $\rho ^{\left( 1/2\right) }$ 
with the probability distribution function
$w\left( \pm1/2,\theta ,\varphi \right) $ of the $1/2$-spin projection on the 
quantization axis, where the angles $0\le \theta \le \pi$,
$0\le \varphi \le 2\pi $ determine the coordinates of the vector normal to 
the surface of the sphere of the unity radius. On the other hand, quantum 
characteristics such as rotational energy levels of molecules and nuclei are 
connected with the tomographic description of the tops' behavior. Therefore 
in this work it will be also discussed the classical-like description of the 
quantum states of a top in terms of the distribution functions. For 
simplicity of presentation, we will consider the example of a symmetric top. 
These results also are discussed in \cite{Safonov1,Safonov2,Andreev}.

\section{Tomography of two-level atom states}

It is a common knowledge (see, for example \cite{Landau!Lifschiz}) that the 
wave function $\psi \left( m\right) $ of the $1/2$-spin particle, which 
can describe the state of a two-level atom and consists of two components 
$\psi \left( 1/2\right) $ and
$\psi \left( -1/2\right) $, can be represented in the form of a spinor
\begin{equation}
\psi =\left( 
\begin{array}{c}
\psi \left( 1/2\right)  \\ 
\psi \left( -1/2\right) 
\end{array}
\right) =\left( 
\begin{array}{c}
a \\ 
b
\end{array}
\right) .
\label{spinor}
\end{equation}
In the case of the pure particle's state, the Hermitian density matrix has 
the form of a $2\times 2$ matrix 
\begin{equation}
\rho ^{\left( 1/2\right) }=\psi \psi ^{\dagger }=\left( 
\begin{array}{cc}
\mid a\mid ^2 & ab^{*} \\ 
ba^{*} & \mid b\mid ^2
\end{array}
\right) ,
\label{m!d}
\end{equation}
where the diagonal matrix elements satisfy the normalization condition
$\mid a\mid ^2+\mid b\mid ^2=1$.
Any rotation in the three-dimensional space, determined by the Euler angles
$\left( \varphi ,\theta ,\psi \right) $ varying in the domain 
$0\le \varphi \le 2\pi $, $0\le \theta \le \pi $, $0\le \phi \le 2\pi $, is 
associated with a $2\times 2$ unitary matrix
\begin{equation}
u\left( \varphi ,\theta ,\psi \right) =\left( 
\begin{array}{cc}
\cos \frac \theta 2\exp \left[\frac{i\left( \varphi +\psi \right) }{2}\right] 
& \sin \frac
\theta 2\exp \left[-\frac{i\left( \varphi -\psi \right) }2\right] \\ 
-\sin \frac \theta 2\exp \left[\frac{i\left( \varphi -\psi \right) }2\right] 
& \cos \frac
\theta 2\exp \left[-\frac{i\left( \varphi +\psi \right) }2\right]
\end{array}
\right).
\label{rotation} 
\end{equation}
Under rotation, the Hermitian density matrix $\rho ^{\left( 1/2\right) }$ 
is transformed as
\begin{equation}
\rho ^{\left( 1/2\right) }\rightarrow \rho ^{\left( 1/2\right) }\left( 
u\right) =u\rho ^{\left( 1/2\right) }u^\dagger ,
\end{equation}
where the $2\times 2$ unitary matrix $u$ is a representation of the angle
that describes the matrix elements of the matrix $u$. For the diagonal elements
of the Hermitian density matrix 
$\rho ^{\left( 1/2\right) }\left( u\right) $ we get
\begin{equation}
\rho ^{\left( 1/2\right) }_{ii}\left( u\right) =\sum_{s=-1/2}^{1/2}\sum%
_{m=-1/2}^{1/2}D_{is}^{\left( 1/2\right) }\left( u\right) 
\rho ^{\left( 1/2\right) }_{sm}D_{im}^{\left( 1/2\right) *}\left( u\right) .
\label{diagon!elem} 
\end{equation}
Here $D_{is}^{\left( 1/2\right) }$ is the Wigner D-function and $i=-1/2,~1/2$.
The diagonal elements of the density matrix of the quantum state take
nonnegative values, and their sum is equal to unity. The physical meaning
of these elements is that they are a probability to measure the value of the
spin projection on the fixed axis in a space. Then, we introduce the notation
\begin{equation}
\rho ^{\left( 1/2\right) }_{ii}\left( u\right) =w\left( i,u\right) ,
\label{den!mar}
\end{equation}
where the function $w\left( i,u\right) $ is the marginal distribution, i.e.,
the probability of finding the spin projection $i$ on the fixed axis in the space 
for the $1/2$-spin particle. From Eq. (\ref{den!mar}), we see that the marginal
distribution also depends on the Euler angles $\varphi ,\theta ,\psi $ as 
parameters. This distribution function is normalized
for all values of the Euler angles. From the structure of Eq. (\ref{diagon!elem}),
it follows that the marginal distribution $w\left( i,u\right) $ depends only
on two Euler angles, and does not depend on the angle $\psi $ of the
rotation. In fact, according to Eq. (\ref{diagon!elem}), one can obtain 
the marginal distribution $w\left( i,u\right) $ of the $1/2$-spin particle 
in the following form
\begin{equation}
w\left( 1/2,u\right) =\cos ^2\frac \theta 2\mid a\mid ^2+\frac{\sin \theta }%
2e^{i\varphi }ab^{*}+\frac{\sin \theta }2e^{-i\varphi }ba^{*}+\sin ^2\frac
\theta 2\mid b\mid ^2
\label{mar!dist!spinor!pos}
\end{equation}
and
\begin{equation}
w\left( -1/2,u\right) =\sin ^2\frac \theta 2\mid a\mid ^2-\frac{\sin \theta }%
2e^{i\varphi }ab^{*}-\frac{\sin \theta }2e^{-i\varphi }ba^{*}+\cos ^2\frac
\theta 2\mid b\mid ^2.
\label{mar!dist!spinor!op}
\end{equation}
Hence, the family of the probability distribution functions of the $1/2$-spin
projection is parametrized by the point's coordinates $\theta $, $\varphi $ 
on the sphere of unity radius. This parametrization coincides with the
physical meaning of the marginal distribution in the sense that the 
distribution function $w\left( i,u\right) $ is the probability to observe the
spin projection $i$ if we measure this spin projection on the quantization
axis which is parallel to the vector normal to the surface of the sphere of
the unity radius in the point with the coordinates $\theta $ and $\varphi $.
If we know the positive, normalized marginal distribution
$w\left( i,u\right) $, then, as it was shown in \cite{Dodonov,Olga1}, the 
matrix elements $\rho ^{\left( j\right) }_{mm^\prime } $ can be calculated with
the help of the measurable marginal distribution $w\left( i,u\right) $  
of the particle with an arbitrary spin $j$ and the values of indices
$i=-j,-j+1,\dots ,j$ by means of the relation
\begin{eqnarray}
\left( -1\right) ^{m^{\prime }}\rho _{mm^{\prime }}^{\left( j\right) }
&=&\sum_{k=0}^{2j}\sum_{l=-k}^k\left( 2k+1\right) ^2\sum_{i=-j}^j\left(
-1\right) ^i  \nonumber \\
&&\ \otimes \int w\left( i,u\right) D_{0l}^k\left( u\right) \frac{d\Omega }{%
8\pi ^2}\left( 
\begin{array}{ccc}
j & j & k \\ 
i & -i & 0
\end{array}
\right) \left( 
\begin{array}{ccc}
j & j & k \\ 
m & -m^{\prime } & l
\end{array}
\right) ,
\label{fig1}
\end{eqnarray}
where $m,~m^\prime =-j,-j+1,\dots ,j$ and the integration leads over the 
rotation angles 
$\varphi ,\theta ,\psi $
\begin{equation}
\int d\Omega =\int\limits_0^{2\pi }d\varphi \int\limits_0^{2\pi }d\psi 
\int\limits_0^\pi \sin
\theta d\theta .
\label{integration}
\end{equation}
In the case of the $1/2$-spin (i.e., $j=1/2$), Eq. (\ref{fig1}) reads 
\begin{eqnarray}
\rho _{mm^{\prime }}^{\left( 1/2\right) } &=&\left( -1\right) ^{-m^{\prime
}}\sum_{k=0}^1\sum_{l=-k}^k\left( 2k+1\right) ^2\sum_{i=-1/2}^{1/2}\left(
-1\right) ^i  \nonumber \\
&&\ \otimes \int w\left( i,u\right) D_{0l}^k\left( u\right) \frac{d\Omega }{%
8\pi ^2}\left( 
\begin{array}{ccc}
\frac 12 & \frac 12 & k \\ 
i & -i & 0
\end{array}
\right) \left( 
\begin{array}{ccc}
\frac 12 & \frac 12 & k \\ 
m & -m^{\prime } & l
\end{array}
\right) ,
\label{always}
\end{eqnarray}
where $m,~m^\prime =-1/2,~1/2 $. 
Calculating all non-zero integrals with D-functions and the marginal 
distributions and taking into account the expressions of 3j-symbols 
(see, for example, \cite{Landau!Lifschiz}),
we can rewrite Eq. (\ref{always}) as follows
\begin{eqnarray}
\rho ^{\left( 1/2\right) }_{mm^{\prime 
}} &=&
\left( -1\right) ^{-m^\prime }\frac i{\sqrt{2}%
}\left( 
\begin{array}{ccc}
\frac 12 & \frac 12 & 0 \\ 
m & -m^{\prime } & 0
\end{array}
\right) \left( \mid a\mid ^2+\mid b\mid ^2\right) \nonumber \\
&&+\frac{3i}{\sqrt{6}}\left( 
\begin{array}{ccc}
\frac 12 & \frac 12 & 1 \\ 
m & -m^{\prime } & 0
\end{array}
\right) \left( \mid a\mid ^2-\mid b\mid ^2\right)   \nonumber \\
&&+i\sqrt{3}\left( 
\begin{array}{ccc}
\frac 12 & \frac 12 & 1 \\ 
m & -m^{\prime } & -1
\end{array}
\right) ab^{*}-i\sqrt{3}\left( 
\begin{array}{ccc}
\frac 12 & \frac 12 & 1 \\ 
m & -m^{\prime } & 1
\end{array}
\right) ba^{*},
\label{reverse!mat!den}
\end{eqnarray}
where $m,~m^\prime =-1/2,~1/2 $, and $i$ is the imaginary unity. 
The substitution of $m,~m^\prime =-1/2,~1/2$ in Eq. (\ref{reverse!mat!den})
leads to the following result:
\begin{equation}
\rho ^{\left( 1/2\right) }_{mm^\prime }=\mid a\mid ^2,~~m=m^\prime =1/2,~~\rho
^{\left( 1/2\right) }_{mm^\prime }=\mid b\mid ^2,~~m=m^\prime =-1/2,
\end{equation}
and in all the other cases
\begin{equation}
\rho ^{\left( 1/2\right) }_{mm^\prime }=ab^*,~~m=-m^\prime =1/2,~~\rho
^{\left( 1/2\right) }_{mm^\prime }=ba^*,~~-m=m^\prime =1/2.
\end{equation}
Then, we checked by the direct calculation that the right-hand side of Eq.
(\ref{always}) is equal to the density matrix (\ref{m!d}).
One can conclude, that given a measurable marginal distribution of a particle,
whose state is described in terms of spinor, one can reconstruct the
state density matrix by means of Eq. (\ref{always}).

\section{Examples of the marginal distribution for the $1/2$ 
and $1$ 
spin states}
\label{example}

The classical-like description of quantum mechanics in terms of the positive,
normalized marginal distribution can be easily understood in the case of the
$1/2$-spin particle. It is a common knowledge \cite{Landau!Lifschiz},
that the spin operator of the $1/2$-spin particle has the form
\begin{equation}
\widehat{s}=\frac 12\widehat{\sigma },
\end{equation}
where the Pauli matrixes are 
\begin{equation}
\widehat{\sigma }_x=\left( 
\begin{array}{cc}
0 & 1 \\ 
1 & 0
\end{array}
\right) ,~~~\widehat{\sigma }_y=\left( 
\begin{array}{cc}
0 & -i \\ 
-i & 0
\end{array}
\right) ,~~~\widehat{\sigma }_z=\left( 
\begin{array}{cc}
1 & 0 \\ 
0 & -1
\end{array}
\right).
\label{Pauly}
\end{equation}
The wave function of the $1/2$-spin particle can be represented in 
the form of spinor (\ref{spinor}). When the direction of the particle's spin
coincided with the positive direction of the $x$ axis in the coordinate
system ($x$, $y$, $z$), the wave function of spinor has the form 
\begin{equation}
\psi _x^{\left( +\right) }=\frac 1{\sqrt{2}}\left( 
\begin{array}{c}
1 \\ 
1
\end{array}
\label{w!fun!x!pos}
\right) .
\end{equation}
And in the case of a contrary direction of the spin with respect to the 
$x$ axis, the wave function reads
\begin{equation}
\psi _x^{\left( -\right) }=\frac 1{\sqrt{2}}\left( 
\begin{array}{c}
1 \\ 
-1
\end{array}
\label{w!fun!x!op}
\right).
\end{equation}
Analogously for the $y$-projection, when the direction of the particle's spin 
is $+1/2$ or $-1/2$ on the $y$ and $z$ axes, we have
\begin{equation}
\psi _y^{\left( +\right) }=\frac 1{\sqrt{2}}\left( 
\begin{array}{c}
1 \\ 
i
\end{array}
\right) ,
\end{equation}
\begin{equation}
\psi _y^{\left( -\right) }=\frac 1{\sqrt{2}}\left( 
\begin{array}{c}
1 \\ 
-i
\end{array}
\right) ,
\end{equation}
and for the $z$-projection
\begin{equation}
\psi _z^{\left( +\right) }=\left( 
\begin{array}{c}
1 \\ 
0
\end{array}
\right) ,
\label{w!fun!z!pos}
\end{equation}
\begin{equation}
\psi _z^{\left( -\right) }=\left( 
\begin{array}{c}
0 \\ 
1
\end{array}
\right) .
\label{w!fun!z!op}
\end{equation}
For all values of the $1/2$-spin directions on the
$x$, $y$ and $z$ axes the density matrices have the following forms
\begin{equation}
\rho _x^{\left( +\right) }=\frac 12\left( 
\begin{array}{cc}
1 & 1 \\ 
1 & 1
\end{array}
\right) ,~~~~\rho _x^{\left( -\right) }=\frac 12\left( 
\begin{array}{cc}
1 & -1 \\ 
-1 & 1
\end{array}
\right),
\end{equation}
\begin{equation}
\rho _y^{\left( +\right) }=\frac 12\left( 
\begin{array}{cc}
1 & -i \\ 
i & 1
\end{array}
\right) ,~~~~\rho _y^{\left( -\right) }=\frac 12\left( 
\begin{array}{cc}
1 & i \\ 
-i & 1
\end{array}
\right),
\end{equation}
and
\begin{equation}
\rho _z^{\left( +\right) }=\left( 
\begin{array}{cc}
1 & 0 \\ 
0 & 0
\end{array}
\right) ,~~~~\rho _z^{\left( -\right) }=\left( 
\begin{array}{cc}
0 & 0 \\ 
0 & 1
\end{array}
\right).
\end{equation}
Applying Eqs. (\ref{mar!dist!spinor!pos}) and (\ref{mar!dist!spinor!op}), we 
obtain the marginal distributions for the different values of the $1/2$-spin
directions on the $x$, $y$ and $z$ axes
\begin{equation}
w_x^{\left( +\right) }\left( +1/2,u\right) =\frac 12\left( 1+\sin 
\theta \cos \varphi \right) ,
~~~~w_x^{\left( +\right) }\left( -1/2,u\right) =\frac 12\left( 1-\sin 
\theta \cos \varphi \right) 
\label{m!dis!x!pos}
\end{equation}
for the positive $x$-projection,
\begin{equation}
w_x^{\left( -\right) }\left( +1/2,u\right) =\frac 12\left( 1-\sin 
\theta \cos \varphi \right) ,
~~~~w_x^{\left( -\right) }\left( -1/2,u\right) =\frac 12\left( 1+\sin 
\theta \cos \varphi \right) 
\label{m!dis!x!op}
\end{equation}
for the negative $x$-projection,
\begin{equation}
w_y^{\left( +\right) }\left( +1/2,u\right) =\frac 12\left( 1+\sin 
\theta \sin \varphi \right) ,
~~~~w_y^{\left( +\right) }\left( -1/2,u\right) =\frac 12\left( 1-\sin 
\theta \sin \varphi \right)
\end{equation}
for the positive $y$-projection,
\begin{equation}
w_y^{\left( -\right) }\left( +1/2,u\right) =\frac 12\left( 1-\sin 
\theta \sin \varphi \right) ,
~~~~w_y^{\left( -\right) }\left( -1/2,u\right) =\frac 12\left( 1+\sin 
\theta \sin \varphi \right) 
\end{equation}
for the negative $y$-projection,
\begin{equation}
w_z^{\left( +\right) }\left( +1/2,u\right) =\cos ^2\frac \theta 2 ,
~~~~w_z^{\left( +\right) }\left( -1/2,u\right) =\sin ^2\frac \theta 2
\label{m!dis!z!pos}
\end{equation}
for the positive $z$-projection and
\begin{equation}
w_z^{\left( -\right) }\left( +1/2,u\right) =\sin ^2\frac \theta 2 ,
~~~~w_z^{\left( -\right) }\left( -1/2,u\right) =\cos ^2\frac \theta 2
\label{m!dis!z!op}
\end{equation}
for the negative $z$-projection. In Figs. 1--2, we plot the marginal 
distributions of the $1/2$-spin particle 
$w_x^{\left( +\right) }\left( \pm1/2,\theta ,\varphi \right) $ the spin
direction of which coincides with the positive direction of the $x$ axis and 
the spin projection onto a fixed axis in space is equal to $\pm 1/2$, as a
function of the angles $\theta $ and $\varphi $.

However, there exist the mixed states of the $1/2$-spin particle
(see, for example, \cite{Landau!Lifschiz}). These states are described only by 
the density matrix
\begin{equation}
\rho _m=\left( 
\begin{array}{cc}
\frac 12+\overline{s}_z & \overline{s}_- \\ 
\overline{s}_+ & \frac 12-\overline{s}_z
\end{array}
\right).
\label{den!matrix!mixed} 
\end{equation}
In Eq. (\ref{den!matrix!mixed}), $\overline{s}_{\pm }$ are determined by the
relation $\overline{s}_{\pm }=\overline{s}_x\pm i\overline{s}_y$,
and $\overline{s}_x$, $\overline{s}_y$ and $\overline{s}_z$ are the mean values
of the $1/2$-spin projections on the $x$, $y$ and $z$ axes, respectively. These
parameters satisfy the condition
$\overline{s}_x^2+\overline{s}_y^2+\overline{s}_z^2\le 1/4$.
A parameter $\mu =Tr\left( \rho ^2_m\right) $ is named ``purity state degree'' 
and depends on the mean values of the spin projection
$\mu =1/2+2\left( \overline{s}_x^2+\overline{s}_y^2+\overline{s}_z^2\right) $.
As in the case of a pure state, using Eq. (\ref{diagon!elem}) for the matrix
elements of the density matrix of a mixed state, we calculate the marginal
distribution of the mixed state of the $1/2$-spin particle,
\begin{equation}
w_m\left( \frac 12,u\right) =\frac 12+\overline{s}_z\cos \theta +\overline{s}%
_x\sin \theta \cos \varphi +\overline{s}_y\sin \theta \sin \varphi 
\label{m!dis!mixed1}
\end{equation}
and
\begin{equation}
w_m\left( -\frac 12,u\right) =\frac 12-\overline{s}_z\cos \theta -\overline{s%
}_x\sin \theta \cos \varphi -\overline{s}_y\sin \theta \sin \varphi .
\label{m!dis!mixed2}
\end{equation}
It is easy to check that the examples of the pure-state marginal distribution 
(\ref{m!dis!x!pos})-(\ref{m!dis!z!op}) are obtained from Eqs. 
(\ref{m!dis!mixed1}), (\ref{m!dis!mixed2}) by the proper choice
of the parameters $\overline{s}_x$, $\overline{s}_y$ and $\overline{s}_z$.

In this section, one can also consider another one important example of the
$1$-spin particle states, the direction of which coincides with 
the positive direction of the $z$ axis. In this particular case, the wave 
function has the form   
\begin{equation}
\psi _z^{\left( 1\right) }=\left( 
\begin{array}{c}
1 \\ 
0 \\ 
0
\end{array}
\right),
\label{w!fun!1}
\end{equation}
and the density matrix, consequently, takes the form
\begin{equation}
\rho ^{\left( 1\right) }=\left( 
\begin{array}{ccc}
1 & 0 & 0 \\ 
0 & 0 & 0 \\ 
0 & 0 & 0
\end{array}
\right).
\label{m!d4}
\end{equation}
The marginal distribution of the spin state can be calculated by the following 
equation (see, for example, \cite{Landau!Lifschiz})
\begin{equation}
\rho _{ii}^{\left( 1\right) }\left( u\right) =D_{si}^1\left( u\right) \rho
_{sm}^{\left( 1\right) }D_{mi}^{1*}\left( u\right),
\end{equation}
where $u$ is determined by Eq. (\ref{integration}). As a result, we obtain
the marginal distribution $w\left( i,\varphi ,\theta ,\psi \right) \equiv 
\rho _{ii}^{\left( 1\right) }\left( u\right) $, 
$\sum _{i=-1}^1 w\left( i,\phi ,\theta ,\varphi \right)=1$, i.e., three
probabilities for the $1$-spin projection onto the $z$ axes ($-1$, $0$, and 
$+1$), which also depend on the rotation angles of the reference frame
$\varphi $, $\theta $, and $\psi $
\begin{equation}
w\left( 1,\theta \right) =\frac{\left( 1+\cos \theta \right) ^2}4,
\label{MD1}
\end{equation}
\begin{equation}
w\left( 0,\theta \right) =\frac{\left( 1-\cos ^2\theta \right) }2,
\end{equation}
\begin{equation}
w\left( -1,\theta \right) =\frac{\left( 1-\cos \theta \right) ^2}4.
\label{MD2}
\end{equation}
A dependence on the $\phi $, $\varphi $ angles drops out. Substituting this
marginal distribution into formula (\ref{fig1}) by means of which matrix 
elements of the density matrix $\rho _{mm^\prime }^{\left( j\right)}$ can be 
reconstructed by the measurable marginal distribution
$w\left( i,u \right) ,~i=-1,0,1$ and 
executing the calculations, we could verify the correction of this formula in
the particular case of the unity spin. In the case $j=1$, this formula 
consists of twenty seven items. Let us consider all values of the Wigner 
D-functions for the various values of their subscripts and superscripts.
Because the marginal distribution $w\left( i,\theta \right),~i=-1,0,1$ takes
only positive values, and because some D-functions contain a complex-valued 
exponential coefficient, integration of the product of the marginal 
distribution and a D-function over the $\phi ,\theta ,\varphi $ rotation 
variables yields zero. Taking this fact into account, Eq. (\ref{fig1}) 
can be simplified, and we arrive at
\begin{eqnarray}
\begin{array}{rcl}
\rho _{mm^{\prime }}^{\left( 1\right) } &=&\frac{\left( -1\right)
^{m^{\prime }}}{8\pi ^2}\biggl[ \sum\limits_{i=-1}^1\int d\Omega \left( 
-1\right) ^iw\left( i,\theta \right) D_{00}^0\left( 
\begin{array}{ccc}
1 & 1 & 0 \\ 
i & -i & 0
\end{array}
\right) \left( 
\begin{array}{ccc}
1 & 1 & 0 \\ 
m & -m^{\prime } & 0
\end{array}
\right)   \\
&&+9\sum\limits_{i=-1}^1\int d\Omega \left( -1\right) ^iw\left( i,\theta 
\right)
D_{00}^1\left( 
\begin{array}{ccc}
1 & 1 & 1 \\ 
i & -i & 0
\end{array}
\right) \left( 
\begin{array}{ccc}
1 & 1 & 1 \\ 
m & -m^{\prime } & 0
\end{array}
\right)   \\
&&+25\sum\limits_{i=-1}^1\int d\Omega \left( -1\right) ^iw\left( i,\theta 
\right)
D_{00}^2\left( 
\begin{array}{ccc}
1 & 1 & 2 \\ 
i & -i & 0
\end{array}
\right) \left( 
\begin{array}{ccc}
1 & 1 & 2 \\ 
m & -m^{\prime } & 0
\end{array}
\right) \biggr] .
\end{array}
\label{matrix!density2}
\end{eqnarray}
Calculating all nonzero integrals and taking into account the expressions 
of the $3j$-symbols from \cite{Landau!Lifschiz} 
we can rewrite Eq. (\ref{matrix!density2}) into the form
\begin{eqnarray}
\begin{array}{rcl}
\rho _{mm^{\prime }}^{\left( 1\right) }&=&\left( -1\right) ^{-m^{\prime
}+1}\biggl[ \frac 1{\sqrt{3}}\left( 
\begin{array}{ccc}
1 & 1 & 0 \\ 
m & -m^{\prime } & 0
\end{array}
\right) \\
&&+\frac 3{\sqrt{6}}\left( 
\begin{array}{ccc}
1 & 1 & 1 \\ 
m & -m^{\prime } & 0
\end{array}
\right) +\frac 5{\sqrt{15}}\left( 
\begin{array}{ccc}
1 & 1 & 2 \\ 
m & -m^{\prime } & 0
\end{array}
\right) \biggr] . 
\end{array}
\end{eqnarray}
For $m\ne m^\prime$ ($m,~m^\prime=-1,~0,~1$), 
$3j$-symbols are equal to zero (see, for example, \cite{Landau!Lifschiz}). 
Evaluating all the remaining $3j$-symbols for $m=m^\prime $, we obtain
\begin{eqnarray}
\rho _{mm^{\prime }}^{\left( 1\right) }=\left( 
\begin{array}{ccc}
1 & 0 & 0 \\ 
0 & 0 & 0 \\ 
0 & 0 & 0
\end{array}
\right) ,~~~~~~m,~m^{\prime }=-1,~0,~1.
\end{eqnarray}
This means that we demonstrated how the initial density matrix (\ref{m!d4}) 
can be recovered, if we know the marginal distribution of the quantum state 
(\ref{MD1})-(\ref{MD2}).%

\section{Tomography of rotational states of molecules and nuclei}

Here we will consider the tomography scheme of rotational states of molecules 
and nuclei using, for simplicity, the model of a symmetric top. It is a common 
knowledge (see, for example,~\cite{Landau!Lifschiz}) that the
Hamiltonian of a symmetric top has the form
\begin{equation}
\widehat{H}=\frac{\hbar ^2}{2I_A}\widehat{J}^2+\frac{\hbar ^2}2\left( \frac
1{I_C}+\frac 1{I_A}\right) \widehat{J_\zeta }^2,
\end{equation}
where $I_A$, $I_C$ are the principal inertia momenta of a top 
(two of the momenta of 
a symmetric top coincide with each other), and $\widehat{\overrightarrow{J}} $ 
is the angular-momentum operator. 
Stationary states of a symmetric top are characterized by three 
quantum numbers: the orbital momentum $j$ and its projections onto the top axis  
($J_\zeta =k, k=-j,-j+1\dots j$) and onto a fixed $z$ axis in space ($J_z=M, 
M=-j,-j+1\dots j$). The symmetric top energy is independent of the 
last quantum number $M$. Let us consider the stationary states of the 
symmetric top with given energy. To do this, we take a subspace 
of the $\left( 2j+1\right) ^2$ dimensions with the fixed $j$ in the state space. 
The wave function of the stationary state of a
symmetric top can be represented in the form
\begin{equation}
\psi _{Mk}^{\left( j\right) }=\langle jMk\mid \psi \rangle =\psi _{jk}^{\left( 
0\right) }D_{kM}^j,
\label{wave!function!top}
\end{equation}
where $D_{kM}^j$ is the Wigner D-function, and $\psi _{jk}^{\left( 0\right)}$ is 
the wave function in the reference frame, which 
comoving
with the
physical system (the top). For a pure state, the density matrix of the symmetric
top is expressed in terms of the wave functions as follows
\begin{equation}
\rho _{MkM^{\prime }k^{\prime }}^{\left( j\right) }=\psi _{Mk}^{\left(
j\right) }\psi _{M^{\prime }k^{\prime }}^{\left( j\right) }.
\end{equation}
Thus, under two 
consecutive rotations (these rotations are determined by the Euler
angles $u\left( \phi ,\theta ,\varphi \right) $ and 
$u^\prime \left(\varphi ^\prime ,\theta ^\prime ,\psi ^\prime \right) $)
the Hermitian density matrix $\rho $ is transformed as
\begin{equation}
\rho \rightarrow \rho ^{\left( j\right) }\left( u,u^{\prime }\right) 
=D^{\left( j\right) }\left( u^{\prime
}\right) D^{\left( j\right) }\left( u\right) \rho ^{\left( j\right) 
}D^{\left( j\right) \dagger }\left( u\right) D^{\left( j\right) \dagger
}\left( u^{\prime }\right),
\label{rotation!m!density}
\end{equation}
where unitary $(2j+1)\times (2j+1)$ matrixes $D\left( u\right) $ and 
$D\left( u^{\prime }\right) $ represent these two rotations through the 
angles $u$ and $u^{\prime }$.
Since the density matrix under consideration depends on four discrete 
indices, it is necessary to do two 
consecutive rotations of the reference frame, which comoving
with the top
to obtain the diagonal elements of the transformed density matrix. 
Taking into account this fact, for the 
diagonal elements of the density matrix  
$\rho ^{\left( j\right) }\left( u,u^{\prime }\right) $ can be reduced to the 
form
\begin{equation}
\rho _{i_1i_2i_1i_2}^{\left( j\right) }\left( u,u^{\prime }\right)
=D_{pi_2}^j\left( u^{\prime }\right) D_{ni_1}^j\left( u\right) \rho
_{npsl}^{\left( j\right) }D_{si_1}^{j*}\left( u\right) D_{li_2}^{j*}\left(
u^{\prime }\right).
\label{diagonal}
\end{equation}
Here $D^j_{ni_1}\left( u\right) $ is the above Wigner D-function and 
$i_1,i_2,n,p,s,l=-j,-j+1,\ldots,j$. In Eq. (\ref{diagonal}), we assume that
summation is performed over the repeated indices $n,\, p,\, s,\, l$, but there 
is no summation over the indices
$i_1,\, i_2$. Below, we will assume that the Hermitian nonnegative density 
matrix $\rho ^{\left( j\right) }_{npsl}$ describes not only the pure state 
of the top (\ref{wave!function!top}) but also an arbitrary mixed state, i.e.  
$Tr{\left( \rho ^{\left( j\right) }\right) }^2<1$.

Let us discuss the problem of reconstructing all matrix elements 
$\rho ^{\left( j\right) }_{npsl}$, if the diagonal elements
$\rho ^j_{i_1i_2i_1i_2}\left( u,u^\prime \right) $ are known. 
To do this, let us 
multiply both sides of Eq. (\ref{diagonal}) by the functions
$D^c_{ab}\left( u\right) $ and $D^f_{de}\left( u^{\prime }\right) $ with
arbitrary upper indices $c,\, f$, and with different angular variables
$u$ and $u^{\prime }$. As a result, we have 
\begin{eqnarray}
\rho _{i_1i_2i_1i_2}^{\left( j\right) }\left( u,u^{\prime }\right)
D_{ab}^c\left( u\right) D_{de}^f\left( u^{\prime }\right) &=&D_{pi_2}^j\left(
u^{\prime }\right) D_{ni_1}^j\left( u\right) \rho _{npsl}^{\left( j\right)
}D_{si_1}^{j*}\left( u\right) D_{li_2}^{j*}\left( u^{\prime }\right) \nonumber \\
&&\otimes D_{ab}^c\left( u\right) D_{de}^f\left( u^{\prime }\right).
\label{times}
\end{eqnarray}
Integrating over the angular variables $u\left( \phi ,\theta ,\varphi \right) $
and $u^\prime \left( \phi ^\prime ,\theta ^\prime ,\varphi ^\prime \right) $,
\begin{equation}
\int d\Omega =%
\frac 18
\int\limits_0^{2\pi} d\phi \int\limits_0^\pi \sin \theta d\theta 
\int\limits_0^{2\pi} d\varphi,
\label{integral}
\end{equation}
\begin{equation}
\int d\Omega ^{\prime }=%
\frac 18
\int\limits_0^{2\pi} d\phi
^{\prime }\int\limits_0^\pi \sin \theta ^{\prime
}d\theta ^{\prime }\int\limits_0^{2\pi} d\varphi ^{\prime },
\end{equation}
using the well-known expression of the integral of the
three D-functions product~\cite{Landau!Lifschiz} and
\begin{equation}
D_{m^{\prime }m}^{j*}\left( u\right) =\left( -1\right) ^{m^{\prime
}-m}D_{-m^{\prime },-m}^j\left( u\right), 
\label{flatly}
\end{equation}
along with the orthogonality properties of the $3j$-symbols, one can express
$\rho ^{\left( j\right) }_{npsl}$ in terms of the diagonal elements  
$\rho ^{\left( j\right) }_{i_1i_2i_1i_2}\left( u,u^\prime \right) $.
As it was mentioned above, the diagonal elements of the density matrix for a 
quantum state take nonnegative values and their sum is equal to unity. Thus,
we introduce the notation
$\rho _{i_1i_2i_1i_2}^{\left( j\right) }\left( u,u^{\prime }\right) =w\left(
i_1,i_2,u,u^{\prime }\right) $,
where the function $w\left( i_1,i_2,u,u^\prime \right) $ is the marginal 
distribution of a symmetric top, i.e. the probability of finding the 
projection $i_1$ of the angular momentum onto a fixed axis in space and the 
projection $i_2$ of the angular momentum onto the top axis, and also this 
probability parametrically depends on Euler angles 
$\phi ,\theta ,\psi ,\phi ^\prime ,\theta ^\prime ,\psi ^\prime $. 
This function is normalized 
$\sum_{i_1,i_2=-j}^jw\left( i_1,i_2,u,u^\prime \right) =1$
for all values of the Euler angles. Suppose we know the positive, normalized 
marginal distribution $w\left( i_1,i_2,u,u^\prime\right) $. Thus,
in view of Eqs. (\ref{times})-(\ref{flatly}) the matrix elements
$\rho ^{\left( j\right) }_{m_1m_2m^\prime _1m^\prime _2}$  can be expressed in
terms of the measurable marginal distribution 
$w\left( i_1,i_2,u,u^\prime \right) $ with $i_1,i_2=-j,-j+1,\ldots,j$ by using
the relationship
\begin{eqnarray}
\rho _{m_1m_2m_1^{\prime
}m_2^{\prime }}^{\left( j\right) } &=&\sum\limits_{k_1=0}^{2j}%
\sum\limits_{l_1=-k_1}^{k_1}\left( 2k_1+1\right) ^2
\sum\limits_{k_2=0}^{2j} \sum\limits_{l_2=-k_2}^{k_2}\left( 2k_2+1\right) ^2
\nonumber \\  
&&\otimes \sum\limits_{i_1=-j}^{j}%
\sum\limits_{i_2=-j}^{j}\left( -1\right) ^{i_1+i_2-m_1^{\prime }-m_2^{\prime 
}}\int \int w\left( i_1,i_2,u,u^{\prime
}\right) D_{0l_1}^{k_1}\left( u\right) \frac{d\Omega }{8\pi ^2} \nonumber \\ 
&&\otimes D_{0l_2}^{k_2}\left( u^{\prime }\right) \frac{d\Omega ^{\prime 
}}{8\pi ^2}
\left( 
\begin{array}{ccc}
j & j & k_1 \\ 
i_1 & -i_1 & 0
\end{array}
\right) \left( 
\begin{array}{ccc}
j & j & k_1 \\ 
m_1 & -m_1^{\prime } & l_1
\end{array}
\right) \nonumber \\
&&\otimes \left( 
\begin{array}{ccc}
j & j & k_2 \\ 
i_2 & -i_2 & 0
\end{array}
\right) \left( 
\begin{array}{ccc}
j & j & k_2 \\ 
m_2 & -m_2^{\prime } & l_2
\end{array}
\right).
\label{matrix!density} 
\end{eqnarray}
Here $m_1,m_2,m^\prime _1,m^\prime _2=-j,-j+1,\ldots,j$. One can conclude that 
given a measurable marginal distribution of the symmetric top, one can 
reconstruct the stationary-state density matrix of the symmetric top by
means of Eq. (\ref{matrix!density}).

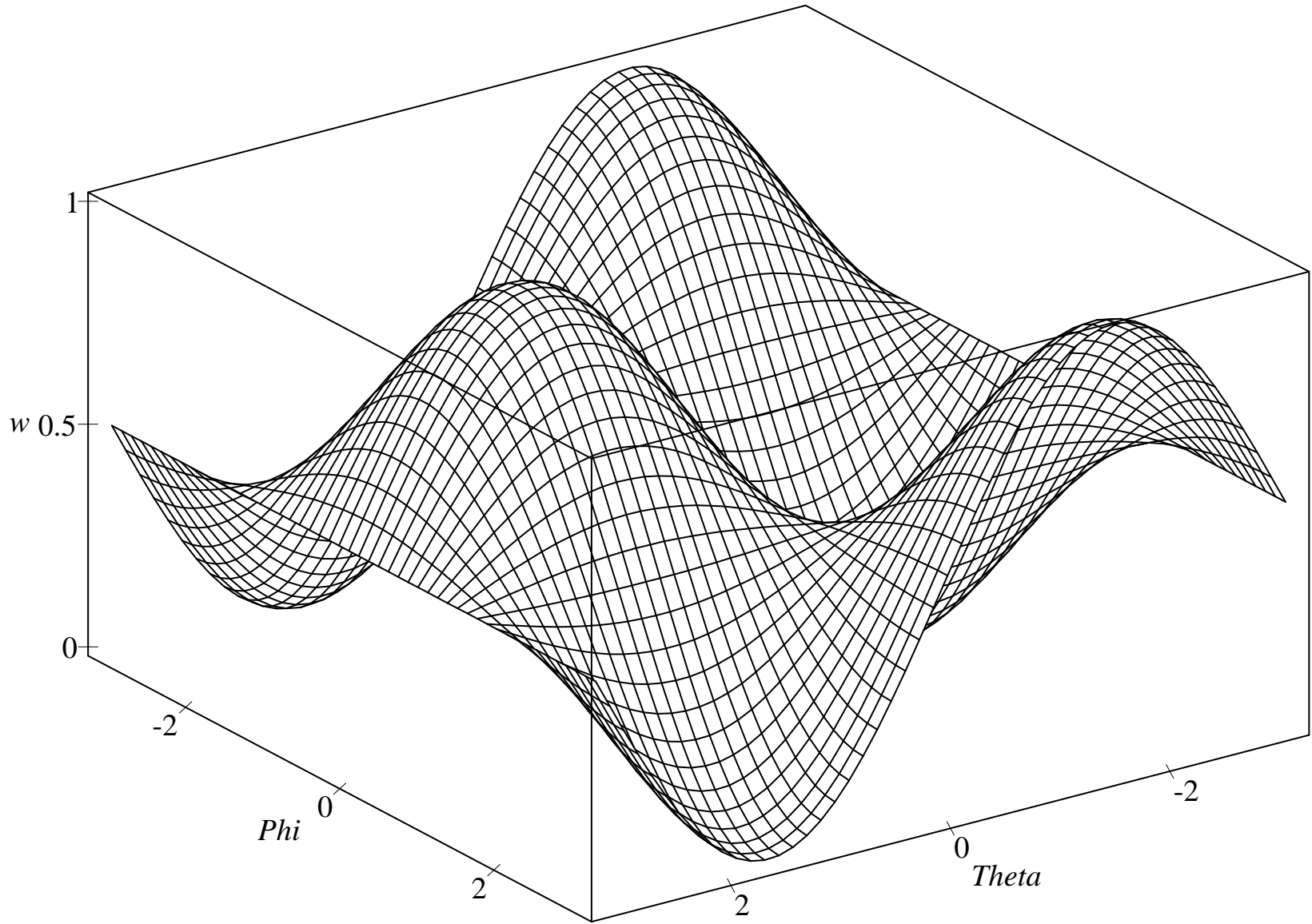
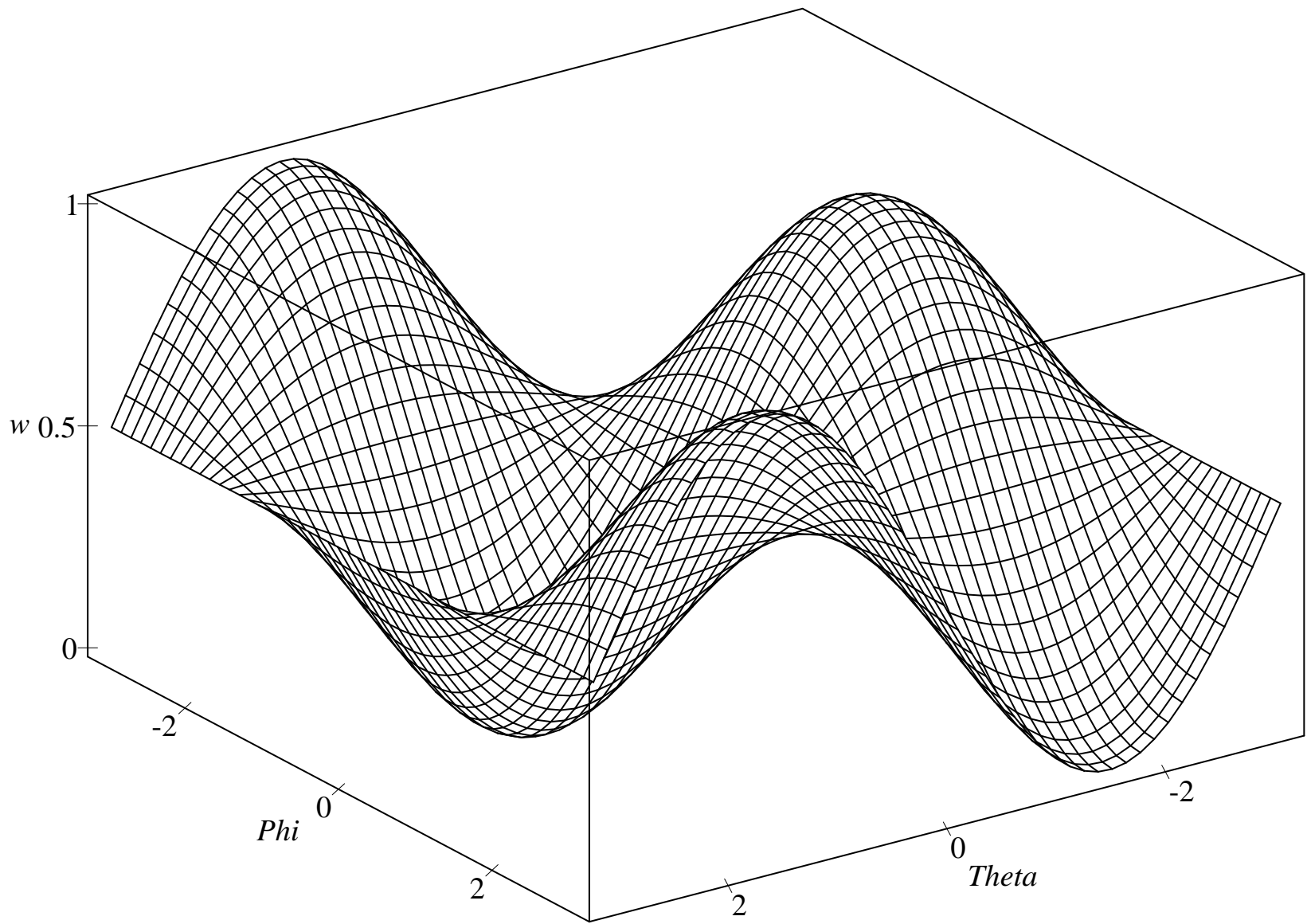
\end{document}